\documentclass{JHEP3}

\input{epsf}
\usepackage{epsfig}
\def\G#1#2 { {\displaystyle \Gamma{ #1 \brack #2  } }  }
\def\bra{\langle}
\def\ket{\rangle}
\def\nn               {  \nonumber  }
\def\ep{\epsilon}

\def\period     { \, . }
\def\comma      { \, , }

\def\F  { {}_2F_1 }
\def\eqb         {  \begin{eqnarray}  }
\def\eqe           {  \end{eqnarray}  }
\def\nn               {  \nonumber  }
\def\lbb     {\left[ }
\def\rbb      {\right] }
\def\o{\over}

\def\re { \mbox{Re} \, }
\def\im { \mbox{Im} \, }
\def\lb       {\left( }
\def\rb       {\right) }

\preprint{MIT-CTP-3308 \\
          UTHEP-463}

 \title{Massless BTZ black holes in minisuperspace}
 \author{Yuji Satoh \\ Institute of Physics \\
University of Tsukuba \\
Tsukuba, Ibaraki 305-8571 \\ Japan 
\\ \email{ysatoh@het.ph.tsukuba.ac.jp}}
\author{ 
Jan Troost\\      Center for Theoretical Physics \\  MIT \\
    77 Mass Ave \\ Cambridge, MA 02139 \\ USA \\
\email{troost@mit.edu}
   }

 \abstract{We study aspects of the propagation of strings on BTZ black holes. 
After performing a careful
analysis of the global spacetime structure of generic BTZ black holes, and its relation
to the geometry of the $SL(2,R)$ group manifold,
we focus on the simplest case of the massless BTZ black hole. We study the $SL(2,R)$ Wess-Zumino-Witten
model in the worldsheet minisuperspace limit, taking into account  special features associated  to the
Lorentzian signature of spacetime. We analyse the two- and three-point functions in the pointparticle limit.
 To lay bare the 
underlying group structure of the correlation functions, we derive new results on Clebsch-Gordan coefficients
for $SL(2,R)$ in a parabolic basis.
 We comment on the application of our results to string theory in
singular time-dependent orbifolds, and to a Lorentzian version 
of the $AdS/CFT$ correspondence.}

\begin{document}
\section{Introduction}
The $AdS/CFT$ correspondence gave rise to the hope that we might understand quantum gravity in black hole
backgrounds better through their field theoretic holographic dual. Since string theory is a consistent
theory of quantum gravity, we can study strings on black hole backgrounds that are asymptotically AdS to try and
realize that hope. In particular, it is known that BTZ-black holes, which are asymptotically $AdS_3$,
can be thought of as orbifolded WZW models. They are stringy backgrounds in which the holographic duality
could be under sufficient control to understand features of strings on singular backgrounds. That's a first
reason to study these backgrounds in more detail.

Recently, singular time-dependent orbifold backgrounds of string theory have been investigated 
(see e.g.\cite{Balasubramanian:2002ry}\cite{Cornalba:2002fi}\cite{Liu:2002ft}
and references thereto).
 It
was argued that, close to the singularity (and in the limit of vanishing cosmological constant), 
a massless BTZ black hole resembles  a parabolic
orbifold of Minkowski space \cite{Cornalba:2002fi}\cite{Simon:2002ma}.
To shed further light on whether
string theory can deal with the parabolic singularity or not 
(see e.g.\cite{Lawrence:2002aj}\cite{Liu:2002kb}\cite{Fabinger:2002kr}\cite{Horowitz:2002mw}), 
it will be very interesting to see whether the
WZW model of the massless BTZ black hole is under good control. Moreover, since it is asymptotically
$AdS_3$, it can perhaps provide us with a holographic interpretation of the troublesome features of
strings on these singular backgrounds. 

One more reason to believe that the massless BTZ black hole background is important in this context, is
the fact that it is supposed to describe the same boundary $CFT_2$ as the $AdS_3$ background, with 
different boundary conditions on the fermions \cite{Coussaert:1994jp}\cite{Maldacena:1998bw}.
 When the boundary $CFT_2$ has $N=2$ supersymmetry
in two dimensions, the boundary field theories are related by spectral flow. That
suggests that at least the boundary conformal field theory corresponding to the massless BTZ black
hole background is well-defined (since the boundary conformal field theory dual to $AdS_3$ backgrounds
is). Thus, a more careful study of this background and its holographic 
properties should shed light on the nature of parabolic orbifolds and their stability.

A formal reason to believe that the orbifolded WZW theories might be manageable, is the underlying group
structure. Therefore, it is important to further analyse the connection between the
structure of the (extended) BTZ spacetimes, and the underlying group manifold and we do so in
section \ref{global}.
In section \ref{masslesssection},
 we concentrate on the massless BTZ black hole, with the motivations above in mind,
and because it is the example with the simplest
global structure.
It is a background that is on the verge of developing a massive black hole with a non-trivial 
horizon, and as such seems very interesting to get to grips with.

In the course of our work we will develop some techniques that should be useful to better
understand typically Lorentzian aspects of the $AdS/CFT$ correspondence. 

\section{Global properties}
\label{global}
In this section we review some of the global properties of the BTZ black hole backgrounds
\cite{Banados:1992wn}\cite{Banados:1993gq} and stress some features that are important
from a group theoretic perspective. That is crucial to us since we would like to study
the Wess-Zumino-Witten model in the BTZ black hole background. 
Most of the properties that
we list are known (see e.g.\cite{Maldacena:1998bw} \cite{Horowitz:1993jc}-\cite{Martinec:2002xq}), 
but some are not, and they are needed for a good understanding of
our treatment of the massless black hole background.  

First of all, it is important to realize that the energy and angular momentum operator
in BTZ black hole backgrounds generate two one-parameter subgroups of $SL(2,R) \times
SL(2,R)$ whose character depends on the particular choice of background. The one-parameter
subgroups are elliptic (two-dimensional rotations) for ordinary $AdS_3$, they
are hyperbolic (two-dimensional boosts) for the generic BTZ black hole, parabolic
(shifts) for the massless BTZ black hole and one hyperbolic and one parabolic 
one-parameter subgroup for the extremal BTZ black hole. Those facts suggest a preferred
choice of coordinates and a preferred basis for the $SL(2,R)$ group manifold. We make 
use of those preferred coordinate choices in the following. 

In \cite{Hemming:2002kd} a nice analysis of the structure of the boundary for
a generic BTZ black hole was presented, and its consequences for the structure
of the boundary conformal field theory were analysed. The analysis made use
of the natural group parametrisation. We want to analyze  similar,
but much simpler properties of the massless BTZ black hole background. To that
end, it will be useful to describe some global features of the BTZ spacetimes,
and their interpretation in terms of the $SL(2,R)$ group manifold.

Since the main part of our paper is on the massless BTZ black hole, the reader
can skip to subsection \ref{massless} on that particular case
 when she is not interested in the following
generalities.

\subsection{$AdS_3$}
For the well-studied  $AdS_3$ background, we choose global coordinates.
To facilitate the diagonalisation of
 the action of elliptic subgroups on kernels (i.e. matrix
elements) of
$SL(2,R)$ representations, we parametrise the group element in the following way:
\begin{eqnarray}
g &=& e^{ \frac{i}{2} (t+\phi) \sigma^2} e^{\rho \sigma^3}  e^{ \frac{i}{2} (t-\phi)\sigma^2}.
\end{eqnarray}
That is the appropriate parametrisation of the $SL(2,R)$ WZW model for the description
of string propagation on the covering, namely $AdS_3$. In $SL(2,R)$ there is a closed timelike
curve, which gives rise to a non-trivial first homotopoy group $\Pi_1(SL(2,R))=Z$.
After unwinding the time direction $t$, we find the 
$AdS_3$ manifold in global coordinates:
\begin{eqnarray}
ds^2 &=& -\cosh^2 \rho dt^2 + d \rho^2 + \sinh^2 \rho d \phi^2.
\end{eqnarray}
Clearly, the boundary of the manifold is
$R \times S^1$ and it is parametrized by $(t, \phi)$ at $\rho \rightarrow \infty $. The $AdS_3/CFT_2$ 
correspondence has been analysed in detail in this case
(see e.g. \cite{Giveon:1998ns}\cite{deBoer:1998pp}\cite{Kutasov:1999xu}\cite{Maldacena:2001km}).
\subsection{Generic BTZ black hole}
The energy and angular momentum operators generate two hyperbolic one-parameter subgroups in
$SL(2,R)\times SL(2,R)$ in the generic BTZ background.
 A parametrisation appropriate for diagonalising  these
operators is:\footnote{
Throughout, we ignore the extra patches corresponding to matrices with a zero.
It would be interesting to clarify whether they play a role in holography.}
\begin{eqnarray}
g &=& e^{\frac{u}{2} \sigma^3} (-1)^{\epsilon_1} (i \sigma^2)^{\epsilon_2} p_i
 e^{\frac{v}{2} \sigma^3} \label{genpara}
\end{eqnarray}
where  $p_i$ takes one of the forms
\begin{eqnarray}
p_1 &=& e^{\rho_1 \sigma^1} \nonumber \\
p_2 &=& e^{i \rho_2 \sigma^2} \label{ps}
\end{eqnarray}
where $\rho_1 \in {]} - \infty, \infty {[}$ and
$\rho_2 \in {[} - \frac{\pi}{4},  \frac{\pi}{4} {[}$. 
The structure of the boundary in this case has been analysed in some detail
in \cite{Hemming:2002kd}. We have little to add to their analysis, but we want
to stress some of the global features of the BTZ spacetime, and how they relate
to the WZW theory.

\subsubsection*{Discrete symmetries}
First of all, we want to point out a few discrete symmetries of the WZW model
that will turn out to be useful \cite{Natsuume:1996ij}.
The following operations are symmetries of the CFT:
the reflection $R: g \rightarrow -g$ and Bargmann's automorphism $B$ that 
amounts to conjugation by $\sigma_3$. It acts on a group element
$g= \left( \begin{array}{cc} a & b \\ c & d \end{array} \right)$ as
$g \rightarrow  \left( \begin{array}{cc} a & -b \\ -c & d \end{array} \right) $. Note that $\sigma_3$ is not 
an element of the group $SL(2,R)$ such that this is an outer automorphism.

Let us now illustrate the action of these symmetries using the (un)extended BTZ spacetime.
This will turn out not to be very fruitful, but will serve as a useful warm-up for what
follows.
By using the parametrisation introduced in (\ref{genpara}), 
we can carve up the group
manifold in 16 regions. First we distinguish eight bigger regions $\pm D_i$ 
where $+D_1$ corresponds to the choice $\epsilon_1=0=\epsilon_2$ and $p_1$
in (\ref{genpara})
and similarly for $i=2$ and $p_2$. The two regions $+D_{3,4}$ similarly correspond to 
the choice $\epsilon_2=1$, $\epsilon_1=0$ and $p_{1,2}$. When $\epsilon_1=1$ 
we denote
the regions $-D_i$ (with the same conventions as before for the index $i$). 
We can moreover split each of the eight regions $\pm D_i$ into two,
creating sixteen patches. We  split each $\pm D_i$ as
$\pm D_i = \pm D_i^+ \cup \pm D_i^-$
according to the sign of $\rho_j$ in (\ref{ps}).

Now, it is easily checked using the description in \cite{Banados:1993gq} of the geometry
that the unextended BTZ spacetime (with mass $M$ and angular momentum $J$),
\begin{eqnarray}
ds^2 & =&  -(-M+\frac{r^2}{l^2}+ \frac{J^2}{4 r^2}) dt^2
+ (-M+\frac{r^2}{l^2}+ \frac{J^2}{4 r^2})^{-1} dr^2
\nonumber \\ & &
+ r^2 (d \phi - \frac{J}{2 r^2} dt)^2 ,
\end{eqnarray}
is equivalent to the group manifold modded out by the
discrete subgroup generated by the two discrete symmetries $R$ and $B$. (A fundamental domain
of the group action is conveniently chosen to be for instance $\cup_i +D^+_i$.)

The extended BTZ spacetime  \cite{Banados:1993gq} 
corresponds to copying the fundamental region an
infinite number of times. It can be thought of as the original group manifold,
where we take the covering with respect to  the closed timelike curve. This essentially reintroduces
the other regions $D_i$, and an infinite number of copies of these regions.
 That agrees (for instance) with the analysis
presented in  \cite{Hemming:2002kd}. 

The discrete symmetries will turn out to be more useful to us in the
 degenerate cases that follow.

\subsection{Extremal black hole}
The story already becomes different for the extremal black hole.
We recall here a few different parametrisations of the extremal black hole
background, firstly to make contact with the group theoretic description,
secondly  because they will turn out to be useful in analysing the 
bulk-to-boundary propagator.

The standard form of the metric for the extremal black hole (where $M=|J|$) is
\begin{eqnarray}
ds^2 &=& - (-M+r^2) dt^2 +
(\frac{M}{2r}- r)^{-2} dr^2 + r^2 d \phi^2 - J d \phi dt.
\end{eqnarray}
In the following we will choose $sign(J)=1$, without loss of generality.
To simplify the description in terms of the natural group coordinates we
perform several coordinate transformations. First, we introduce the
coordinate $U^2 = r^2 - \frac{M}{2}$. In terms of the new coordinates
the metric becomes:
\begin{eqnarray}
ds^2 &=& U^2(- dt^2 + d \phi^2) + \frac{M}{2} (dt - d \phi)^2 + \frac{dU^2}{U^2}.
\end{eqnarray}
These coordinates can be related to standard Poincare coordinates by the transformations:
\begin{eqnarray}
\lambda_+ &=& \frac{1}{2 \pi T_+} e^{2 \pi T_+ (-\phi+t)} \nonumber \\
\lambda_- &=& -\phi -t - \frac{\pi T_+}{U^2} \nonumber \\
z &=& \frac{1}{U} e^{ \pi T_+ (-\phi+t)},
\end{eqnarray}
where the temperature $T_+$ is equal to 
$T_+=\frac{r_+ + r_-}{2 \pi}=\frac{r_+}{\pi}=\frac{1}{\pi}\sqrt{\frac{M}{2}}$ 
for the extremal black hole.
We obtain the metric
\begin{eqnarray}
ds^2 &=& \frac{1}{z^2} (dz^2+d \lambda_+ d \lambda_-).
\end{eqnarray}
Now we can compare to the appropriate
$SL(2,R)$ parametrisation:
\begin{eqnarray}
g &=& (-1)^{l} e^{u L^-} e^{\rho L^+} e^{v \sigma^3},
\end{eqnarray}
where $\sigma^i$ are the Pauli matrices and $L^{\pm}=\frac{1}{2}
(\sigma^1 \pm i \sigma^2)$.
It is natural to first of all mod out by the reflection $R$, such that
we have only one patch, labelled by $l=0$. We can compute the metric
on the group easily.
The coordinates on this patch of the group are related
to Poincare coordinates in a one-to-one fashion:
\begin{eqnarray}
z &=& e^{-v} \nonumber \\
\lambda_+ &=& \rho e^{-2v} \nonumber \\
\lambda_- &=& u
\end{eqnarray}
where $z \in R^+, \lambda_{\pm}, \rho, u,v \in R$.

Checking carefully the coordinate transformations above, it becomes
clear that the region $r>\sqrt{\frac{M}{2}}$ in the standard coordinates
for the extremal BTZ black hole is mapped one-to-one
to the half of the Poincare patch with $\lambda_+>0$. We can obtain
that condition in the WZW model by modding out the model further
by Bargmann's automorphism $B$, which allows us to choose the
sign of $\lambda_+$.

In summary,  the region outside the horizon
of the extremal black hole is mapped one-to-one to the fundamental
domain of the WZW model modded out by $(1,R,B,RB)$.
 Thus we have
a global definition of the BTZ spacetime in terms of the group
manifold.
It can moreover be checked that the region inside the horizon can be 
included in the group theory description by not modding out by $B$.
The extra patch can be mapped to the region inside the horizon using
coordinate transformations very analogous to the ones above.

 Of course we still need to mod out by the appropriate $\phi \rightarrow
\phi+ 2 \pi$ symmetry to obtain the black hole spacetime. It would
be useful to clarify the group theory description of the full extended
extremal BTZ spacetime in \cite{Banados:1993gq}, but we won't need it
here.

\subsection{Massless black hole}
\label{massless}
It turns out that the massless black hole case, with metric
\begin{eqnarray}ds^2 &=& - \frac{r^2}{l^2} dt^2
+ \frac{l^2}{r^2} dr^2 + r^2 d\phi^2
\end{eqnarray}
is the simplest to describe.
We parametrise the group as:
\begin{eqnarray}
g &=& e^{v_l L^-} e^{\rho \sigma^3} (i \sigma^2) e^{-v_r L^-} (-1)^{l}.
\end{eqnarray}
After taking the orbifold by the 
reflection $R$, and restricting to the fundamental
domain labelled by $l=0$, we cover the standard massless black hole
background precisely once, and the range of the radial coordinate is
$r=e^{\rho} \in {]}0, \infty {[}$. 
The boundary at infinity is clearly $R \times S^1$. 

If we would not mod out by $R$, then the
spacetime has a second component. We can interpret the resulting total
group manifold as describing the massless black hole where the radial
coordinate $r$ runs over the
real line $r \in
{]} - \infty , \infty {[}$. In that case, we
have to consider two boundaries of the form $R \times S^1$, one at each
radial infinity. Note that the group generated by $R$ is embedded in
$SL(2,R)$ as $Z_2={\{} 1,-1 {\}} \subset SL(2,R)$. The representations
of $SL(2,R)$ are easily classified as having a definite parity under the
$Z_2$. The above choice of whether to orbifold or not thus corresponds
to the choice of including into the Hilbert space of the model representations
with non-trivial parity or not.

Note that both in the case of the extremal and the massless black hole there
is a group theoretic operation that maps the interior of the black hole to the
exterior.
\section{Massless black hole}
\label{masslesssection}
After the analysis of some global aspects of the family of 
BTZ black holes, we can turn
to the simplest cousin, the massless black hole. Some features of strings on the massless
BTZ black hole background were uncovered already.
In particular its connection to the $AdS_3/CFT_2$ 
correspondence via spectral 
flow in the boundary conformal field theory \cite{Coussaert:1994jp}\cite{Maldacena:1998bw}, 
the construction
of the boundary conformal algebra in terms of worldsheet vertex operators \cite{Troost:2002wk}, 
and the analysis
of the ghost-free spectrum \cite{Bars:1995mf}\cite{Bars:1995cn} for winding strings \cite{Troost:2002wk}
 in the massless black hole background.

But simple aspects of the model were not explored very deeply. In particular, little has been said
about the features of the model associated to the Lorentzian nature 
of spacetime. Although
the analysis of the minisuperspace limit of $SL(2,C)/SU(2)$ (i.e. euclidean $AdS_3$)
was performed in a well-controlled manner \cite{Teschner:1997fv}, 
as was the CFT analysis \cite{Teschner:1997ft}, 
little has been said about the appropriate
Lorentzian analytic continuation. 
(For discussions of the Lorentzian $AdS_3$ in relation to the Euclidean 
one  see 
e.g. \cite{Maldacena:2001km}\cite{Hosomichi:vp}\cite{Satoh:2001bi}.)
Although the Lorentzian two-dimensional CFT is presumably unstable and may be difficult to make sense
of without reference to its euclidean counterpart,
one certainly can analyse Lorentzian aspects of the $AdS/CFT$ correspondence by simply studying a
particle on the group manifold. It has been pointed out several times that it is important to get
a better handle on time-dependent Lorentzian features of the $AdS_3/CFT_2$ correspondence to make progress in 
understanding black hole physics via the corresponding boundary field theory.

By analysing the quantum mechanical particle on the $SL(2,R)$ group manifold, in a way appropriate
for the study of strings on the massless black hole background, we hope to provide good intuition
for the correct Lorentzian interpretation of the appropriate $SL(2,C)/SU(2)$ amplitudes.

\subsection{Model}

The metric for the vacuum black hole with $M=0=J$
is given by 
\begin{eqnarray}ds^2 &=& - \frac{r^2}{l^2} dt^2
+ \frac{l^2}{r^2} dr^2 + r^2 d\phi^2
\end{eqnarray}
with $\phi \in {[}0, 2 \pi {[}$ (and we put $l=1$). 
In terms of the coordinates on the group
we have (in the patch $l=0$)
\begin{eqnarray}
g &=& e^{v_l L^-} e^{\rho \sigma^3} (i \sigma_2)  e^{-v_r L^-}.
 \label{g} 
\end{eqnarray} 
The parametrisation
is related to the previous coordinates by the transformation
\begin{eqnarray}
v_l &=&  \phi +t  \nn \\
v_r &=& \phi - t \nn \\
e^{\rho} &=& r.
\end{eqnarray}
In terms of these coordinates the metric reads:
\begin{eqnarray}
ds^2 &=& e^{2 \rho} dv_l dv_r + d \rho^2,
\end{eqnarray}
which is the invariant metric on the group manifold.
The global angular identification in these coordinates is
$(v_l,v_r,\rho) \equiv (v_l+2 \pi, v_r+ 2 \pi,\rho)$,
which is generated by $(h_l,h_r)=(e^{2 \pi L^-},e^{-2 \pi L^-})$.
\subsection{Minisuperspace}
In this section we discuss the Lagrangian, the Hamiltonian and the
symmetries of the particle model in some detail (see  \cite{Barsnotes}).
The Lagrangian density in the minisuperspace limit is:
\begin{eqnarray}
{\cal L} &=& \frac{1}{2} (\dot{\rho}^2+ e^{2 \rho} \dot{v_l} \dot{v_r}).
\end{eqnarray}
The variables canonically conjugate to the coordinates are:
\begin{eqnarray}
p_l &=& \frac{1}{2} e^{2 \rho} \dot{v_r} \nonumber \\
p_r &=& \frac{1}{2} e^{2 \rho} \dot{v_l} \nonumber \\
p_{\rho} &=& \dot{\rho}.
\end{eqnarray}
The classical Hamiltonian is:
\begin{eqnarray}
{\cal H} &=& \frac{p_{\rho}^2}{2} + 2 e^{-2 \rho} p_l p_r = \frac{p^{\mu}p_{\mu}}{2}.
\end{eqnarray}
To quantise the model, we impose the commutation relations:
\begin{eqnarray}
{[} \rho, p_{\rho} {]} &=&i \nonumber \\
{[} v_l, p_{l} {]} &=&i \nonumber \\
{[} v_r, p_{r} {]} &=&i.
\end{eqnarray}
The symmetry of the model is $SL(2,R) \times SL(2,R)$, which will be broken
to $R \times U(1)$ (i.e. time translations and rotations)
by the global angular identification.
We can define the currents as $J^i_L =k \, Tr(L^i \partial_{+}g g^{-1})$
for the leftmovers, and
$J^i_R =-k \, Tr(L^i g^{-1} \partial_{-}g )$ for the rightmovers,
where $L^3=\frac{i}{2} \sigma^3$.
The leftmoving currents can be computed to be:
\begin{eqnarray}
J^3_L &=& ik ( \partial_{\tau} \rho- e^{2 \rho} v_l 
\partial_{\tau} v_r) \nonumber \\
J^-_L &=& k e^{2 \rho} \partial_{\tau} v_r \nonumber \\
J^+_L &=& k( \partial_{\tau} v_l+ 2 v_l \partial_{\tau} \rho- v_l^2 e^{2 \rho} 
\partial_{\tau} v_r)
\end{eqnarray}
and the rightmoving ones are:
\begin{eqnarray}
J^3_R &=& ik ( \partial_{\tau} \rho- e^{2 \rho} v_r \partial_{\tau} v_l)
 \nonumber \\
J^-_R &=& k e^{2 \rho} \partial_{\tau} v_l \nonumber \\
J^+_R &=& k (\partial_{\tau} v_r+ 2 v_r \partial_- 
\rho - v_r^2 e^{2 \rho} \partial_{\tau} v_l).
\end{eqnarray}
In terms of the canonical variables, we can rewrite the currents as:
\begin{eqnarray}
J^-_L &=& 2k p_l \nonumber \\
J^3_L &=& ik (p_{\rho}- ( v_l p_l+p_l v_l)) \nonumber \\
J^+_L &=& 2k ( e^{-2 \rho} p_r + v_l p_{\rho}-v_l p_l v_l)
\end{eqnarray}
where we chose the ordering  as to have (anti)-hermitian 
currents (and we could have added normal ordering constants).
We have similar formulas for the right-movers.
With these conventions, the commutation relations between the
currents are:
\begin{eqnarray}
{[} J^3_L, J^{\mp}_L {]} &=& \pm 2k  J^{\mp}_L \nonumber \\
{[} J^-_L, J^+_L {]} &=& -4 k J^3_L,
\end{eqnarray}
which indeed generates an $sl(2,R)$ algebra.
The quadratic Casimir $c_2$ is related to the 
quantum mechanical Hamiltonian $H$ for a particle on a group manifold. We have:
\begin{eqnarray}
c_2 &=& -\frac{H}{2} - \frac{1}{4}.
\end{eqnarray}
\subsection{Hilbert space}
The first slightly trickier issue to tackle is the choice of Hilbert space for
the particle on the group manifold. We propose to take
 the Hilbert space to consist of the 
quadratically integrable functions on the group manifold
$SL(2,R)$, and to mod it out by the global angular identification. In a string
theoretic context, that gives rise to twisted states, namely
the winding states discussed in detail in this context in \cite{Troost:2002wk}
following e.g. \cite{Maldacena:2001hw}\cite{Maldacena:2001kv}.
For the string spectrum for $SL(2,R)$, see also \cite{Satoh:2001bi}.

The  (scalar) particle wavefunction in a second quantisation satisfies
the Klein-Gordon equation.  We can classify the solutions using the two
commuting conserved quantities, $p_l$ and $p_r$, which correspond to 
the difference and the sum of the energy $E$ and the angular momentum $L$.
The Klein-Gordon equation in our curved background is:
\begin{eqnarray}
\Delta \Phi (v_l,v_r, \rho) &=& \tau (\tau+1) \Phi,
\end{eqnarray}
where we used the standard form for the quadratic Casimir $c_2=\tau(\tau+1)$.
The equation has solutions:
\begin{eqnarray}
\Phi &=& e^{-i \lambda v_l} e^{i \mu v_r} u(\rho) \nn \\
     &=& e^{-i Et} e^{i L \phi} u(\rho)
\end{eqnarray}
where the energy $E$ is given by $E=\lambda+\mu$ and the angular momentum by
$L=\mu-\lambda$.
The function of the radial variable, 
$u(\rho)$, can be expressed in terms of
a linear combination of Bessel functions.

\subsection*{$E^2>L^2$}
For $4 \lambda\mu= E^2-L^2>0$, we find the solutions:
\begin{eqnarray}
\Phi &=& e^{-i Et} e^{i L \phi} e^{-\rho}
(c_1 J_{2 \tau+1} (\sqrt{4 \lambda \mu} e^{-\rho})
+c_2 J_{-(2 \tau+1)} (\sqrt{4 \lambda \mu} e^{-\rho})).
\end{eqnarray}
For integer  values of the index, we can express the second
linearly independent solution in terms of $Y_{2 \tau+1}$.
\subsection*{$E^2<L^2$}
For $4 \lambda\mu= E^2-L^2<0$, we find the solutions:
\begin{eqnarray}
\Phi &=& e^{-i Et} e^{i L \phi} e^{-\rho}
(c_1 I_{2 \tau+1} (\sqrt{-4 \lambda \mu} e^{-\rho})
+c_2 I_{-(2 \tau+1)} (\sqrt{-4 \lambda \mu} e^{-\rho})).
\end{eqnarray}
For integer values of the index, we need $K_{2 \tau+1}$ 
to find a second linearly independent solution.\footnote{Note
that we don't treat the special case $E^2=L^2$ here, although is has
interesting special properties (see e.g.\cite{Troost:2002wk}) related to 
supersymmetry.}
\subsection{Normalisability and unitarity}
If we restrict to unitary representations of $SL(2,R)$, the possible 
values for $\tau$ are $\tau \in \frac{1}{2} Z$ or 
$-1<\tau<0$ or $\tau=-\frac{1}{2}+is$ where $s\in R$.
When we further restrict our Hilbert space to the quadratically
integrable functions\footnote{
By square integrable functions, we mean the functions
 square integrable against the
group measure $dg = \frac{1}{4 \pi^2} d \phi d t d \rho e^{2 \rho}$,
where we integrate $\rho,t$ over the real line and $\phi$ becomes
an angular variable after the compactification.},
 we only need the principal
continuous ($\tau=-\frac{1}{2}+is$) and discrete 
($\tau \in \frac{1}{2} Z$) representations. 

It can be shown that the space of quadratically integrable functions can be
decomposed into an integral over continuous representations (with two possible
parities) and discrete representations with highest or lowest weight (\cite{VK} p. 488).
For those representations,
the behavior of the particle wavefunctions at infinity is like plane
waves (continuous representations, delta-function normalisable)
or exponential fall-off (discrete representations).\footnote{
There are two linearly independent solutions to the Klein-Gordon
equation. For the discrete series, only one, which falls off exponentially, 
appears as a (normalisable) matrix element. The other solution, which exponentially increases,
is crucial for $AdS/CFT$ though, as it provides a non-fluctuating background,
or non-trivial source term on the boundary \cite{Balasubramanian:1998sn}. }

We return to a more detailed description of this decomposition when
discussing two-point functions in the minisuperspace limit.

\subsection{Kernels}
The wavefunctions for the scalar particle can also be expressed as matrix elements,
or kernels,
of the representations of $SL(2,R)$. Indeed,
the kernels $K^{33}$ of $SL(2,R)$ expressed in a parabolic basis (i.e. diagonalising
$p_r$ and $p_l$),
are eigenfunctions of the
Laplacian with eigenvalue $\tau(\tau+1)$, and represent wavefunctions
of scalar fields in a momentum basis. When normalized properly, they are given by
\cite{VK}:
\begin{eqnarray}
K^{33}(\lambda, \mu; \chi; g) &=& (-1)^{2 \epsilon \delta}
\frac{e^{i \pi \epsilon}}{2 \sin(\pi (\tau+\epsilon+\frac{1}{2}))}
(\frac{\mu}{\lambda})^{\tau+\frac{1}{2}}
e^{-i (\lambda+\mu)t} e^{i(\mu-\lambda)\phi} e^{-\rho}
\nonumber \\
& & 
(J_{-2\tau-1}(\sqrt{4 \lambda \mu} e^{-\rho})
-(-1)^{2 \epsilon}
J_{2 \tau+1}(\sqrt{4 \lambda \mu} e^{-\rho})),
\label{kernels}
\end{eqnarray}
for $\lambda \mu > 0$. 
Here,  $\epsilon = 0,1/2$ denotes the parity of the representation; 
$\delta=\frac{1-\mbox{sign} (\lambda)}{2}$; $\chi=(\tau,\epsilon)$ and $g$
is given by (\ref{g}). 
We have moreover:
\begin{eqnarray}
K^{33}(\lambda, \mu; \chi; g) &=& (-1)^{2 \epsilon \delta}
\frac{1}{\pi}
(-\frac{\mu}{\lambda})^{\tau+\frac{1}{2}}
e^{-i (\lambda+\mu)t} e^{i(\mu-\lambda)\phi} e^{-\rho}
(e^{(\tau+\frac{1}{2})\pi i}
+ (-1)^{2 \epsilon} e^{-(\tau+\frac{1}{2})\pi i}
)
\nonumber \\
& & 
K_{2 \tau+1}(\sqrt{-4 \lambda \mu} e^{-\rho}),
\end{eqnarray}
for $\lambda \mu < 0$.  If we include $(-1)^l$ in $g$, the kernels 
are multiplied by $(-1)^{2\epsilon l}$. 

These formulas are valid when $-1< \re(\tau)<0$, but can easily be extended to the
unitary
discrete series. We thus find that the matrix elements of the unitary,
square integrable representation spaces correspond to the particle
wavefunctions that we computed before.
The global identification $\phi \rightarrow \phi+2 \pi$
restricts the choice of angular momentum $L=\mu-\lambda$
to be integer. We now want to turn to computing
overlaps of these wavefunctions.

\subsection*{Remark}
Note that once we properly defined the Hilbert space for the particle, there is
a natural guess for the string spectrum in this background. 
We take the particle
Hilbert space and build affine representations on these primary states 
to obtain
the untwisted sector of the string Hilbert space.
Next, we include winding, twisted sectors along 
the lines proposed in \cite{Troost:2002wk} by analogy 
to the $AdS_3$ case \cite{Maldacena:2001hw}. 

In the massless black hole case,
it turns out that such the twisted states are related to the Hilbert
space proposed
in \cite{Bars:1995mf}\cite{Bars:1995cn} (for the continuous representations). 
Similarly, the inclusion of 
the twisted sectors in the generic case \cite{Hemming:2001we} following
\cite{Maldacena:2001hw} essentially leads to the construction presented
in \cite{Natsuume:1996ij}\cite{Satoh:1997xe}. 
It would be nice to confirm the prediction for the Hilbert space
by computing the one-loop partition function along the lines 
of \cite{Maldacena:2001kv}.

\section{Overlaps}
In the previous section we constructed the particle Hilbert space using results
from $SL(2,R)$ group theory. In this section, we want to compute overlaps
of the constructed wavefunctions.
These are particle approximations (see e.g. \cite{Teschner:1997fv}) to N-point functions in the
CFT that one obtains when putting a string in the massless black hole background.
For convenience we work in momentum space in this section, since that
is where we most easily obtained the wavefunctions.

In our analysis of the two-point functions, we will clarify further the structure of the
Hilbert space. When we come to the three-point functions, 
we will lay bare their $SL(2,R)$ group structure.

We define the N-point function in momentum space
in the minisuperspace limit  to be given by the N-point overlap:
\begin{eqnarray}
\langle \prod_{i=1}^N  K^{33} (\lambda_i,\mu_i;\chi_i;g) \rangle
& = &
\frac{1}{8 \pi^2} \int d v_r \int dv_l
\int_{-\infty}^{+\infty} d \rho \, e^{2 \rho}
\prod_{i=1}^N K^{33} (\lambda_i,\mu_i;\chi_i;g)  \\
& = & 
     \delta\bigl(\Sigma (\mu_i + \lambda_i) \bigr) 
   \delta_{\Sigma (\mu_i -\lambda_i)}
\int_{-\infty}^{+\infty} d \rho \, e^{2 \rho}
\prod_{i=1}^N K^{33} (\lambda_i,\mu_i;\chi_i;\rho). \nn
\end{eqnarray}
This amounts to imposing conservation of energy and angular
momentum,
and evaluating an integral over $N$ Bessel functions 
of argument $a_i e^{-\rho}$ with measure $e^{(2-N)\rho} \, d \rho$,
where $a_i =\sqrt{\pm \lambda_i \mu_i}$. 
For convenience, we have made a summary of some of the integrals of
Bessel functions that we need in calculating two- and three-point
functions in our appendix \ref{integrals}.

\subsection{Two-point function}
In this subsection, we derive the two-point overlaps for the kernels
above. Of course, these are diagonal in the representation index, but
since the proof of this fact entails some non-trivial functional
analysis, we believe it is useful to incorporate it here. We first
discuss the orthogonality of the Bessel-functions, and then discuss 
how the decomposition of our Hilbert space fits into a generalized
theory of Fourier transforms. Then the analysis of the two-point functions becomes
trivial.

\subsection*{Orthogonality of $J_\mu$}
We believe it is interesting to first of all show how the orthogonality relations among Bessel
functions $J_\mu$
can be derived from the integral formula 
\eqb
   (J_\mu, J_\nu) &\equiv & 
   \int_0^\infty {dt \o t} J_\mu (a t) J_\nu(a t) 
   \ = \ \frac{2}{\pi} {1 \o \mu^2 -\nu^2} \sin {\pi \o 2} (\mu -\nu) 
  \comma \label{intJJ}
\eqe
which is valid for 
\eqb
  \re (\mu + \nu ) > 0 \comma \qquad  a > 0 \period
 \label{cond}
\eqe

\subsubsection*{Detailed example}
Let us first consider the orthogonality of $J_{i\rho} \pm J_{-i\rho}$.
To maintain (\ref{cond}), we regularlise it as 
$J_{i\rho+ \ep} \pm J_{-i\rho + \ep}$ with a positive infinitesimal number 
$\ep$. Using
\eqb
   (J_{i\rho+ \ep}, J_{i\lambda+ \ep}) &=& 
    \frac{2i}{\pi} {\sinh \frac{\pi}{2} (\rho-\lambda) \o \rho - \lambda}
    {-1 \o \rho + \lambda - 2i \ep}
  \comma 
\eqe
and similar expressions, one finds that 
\eqb 
  && (J_{i\rho+ \ep} \pm J_{-i\rho+ \ep}, 
   J_{i\lambda+ \ep} \pm J_{-i\lambda+ \ep}) \nn \\
  &&  \qquad = 
  (J_{i\rho+ \ep}, J_{i\lambda+ \ep}) + (J_{-i\rho+ \ep}, J_{-i\lambda+ \ep})
  \pm \lbb (J_{i\rho+ \ep}, J_{-i\lambda+ \ep}) +
   (J_{-i\rho+ \ep}, J_{i\lambda+ \ep})  \rbb \nn \\
  && \qquad =  
   \frac{4}{\pi}{\sinh \frac{\pi}{2} (\rho-\lambda) \o \rho - \lambda}
    {2 \ep \o (\rho + \lambda)^2 + (2\ep)^2}
  \pm 
 \frac{4}{\pi}{\sinh \frac{\pi}{2} (\rho+\lambda) \o \rho + \lambda}
    {2 \ep \o (\rho - \lambda)^2 + (2\ep)^2} \nn \\
  && \qquad \to 
  2 \frac{\sinh \pi \rho}{\rho} \lbb \delta(\rho+\lambda) 
          \pm \delta(\rho-\lambda) \rbb \period \label{JpmJpm}
\eqe
Here we used the relation 
$\delta (x) = \lim_{\ep \to +0} \frac{1}{\pi} {\ep \o x^2 + \ep^2}$ . 
\subsubsection*{Other cases}
Similarly, we find the following orthogonality relations:
\eqb
  (J_{i\rho} +  J_{-i\rho}, J_{2n}) &=& 0 \nonumber \\
  (J_{i\rho} - J_{-i\rho},  J_{2n+1}) &=&  0 \period
\eqe
and
\eqb
  (J_{2n}, J_{2m}) &=&   \frac{1}{4n} \delta_{n,m} \comma \nn \\
  (J_{2n+1}, J_{2m+1}) &=&   \frac{1}{4n+2} \delta_{n,m} \period 
\eqe

\subsubsection*{Expansions of functions}

The above calculation shows that the sets 
$ \{ J_{i\rho} + J_{-i\rho}, J_{2n} \,| \,
 \rho \in R_{\geq 0}, \, n \in Z_{\geq 1} \}$, and 
$ \{ J_{i\rho} - J_{-i\rho}, J_{2n+1} \, | \,  
 \rho \in R_{\geq 0}, \, n \in Z_{\geq 0} \}$ form orthogonal sets of 
functions, respectively. Actually, in \cite{Titchmarsh} it is shown, using
the general theory of eigenfunctions of second order differential operators,
that 
they form a complete set such that we can expand every quadratically
integrable function on the half-line in terms of these orthogonal bases. 
{From} the expression for the kernels (\ref{kernels}), 
it is clear that the two sets are
related to trivial and non-trivial representations of the reflection
symmetry $R$ that we discussed in subsection \ref{massless}. Thus, when we
restrict to studying the unextended massless BTZ black hole, we would take
along only wavefunctions with trivial parity $\epsilon=0$. When we
include the second patch in the group manifold (associated to $l=1$),
we need to add a second component to our Hilbert space, associated to
wavefunctions with non-trivial parity $\epsilon=1$.

We can expand a function $g(z)$ on the half-line  as
 (see also \cite {Titchmarsh})
\eqb
  g(z) &=& \sum_{n=1}^\infty 4n C_n J_{2n}(z)  
  + \int_{0}^\infty  { \rho d\rho \o 2 \sinh \pi \rho} C_\rho 
      \{ J_{i\rho}(z) + J_{-i\rho}(z)\} \comma \nn \\
  |g(z)|^2 &=& (g, g^*) \ = \ 
    \sum_{n=1}^\infty 4n |C_n|^2 
  +  \int_{0}^\infty  { \rho d\rho \o 2 \sinh \pi \rho} |C_\rho|^2
  \comma
\eqe 
where
\eqb
  C_n &=& (g, J_{2n}) \comma \quad C_\rho \ = \ (g, J_{i\rho} + J_{-i\rho})
  \period
\eqe 
For the expansion by the second set of orthonormal functions we have:
\eqb
  g(z) &=& \sum_{n=0}^\infty (4n+2) C_n J_{2n+1}(z)  
  - \int_{0}^\infty  { \rho d\rho \o 2 \sinh \pi \rho} C_\rho 
      \{ J_{i\rho}(z) - J_{-i\rho}(z)\} \comma \nn \\
  |g(z)|^2 &=& (g, g^*) \ = \ 
    \sum_{n=0}^\infty (4n+2) |C_n|^2 
  +  \int_{0}^\infty  { \rho d\rho \o 2 \sinh \pi \rho} |C_\rho|^2
  \comma
\eqe 
where
\eqb
  C_n &=& (g, J_{2n+1}) \comma \quad C_\rho \ = \ (g, J_{i\rho} - J_{-i\rho})
  \period
\eqe

\subsection*{Orthogonality of $K_\mu$}
Similarly to the case of $J_\mu$, we can derive the orthogonality of 
$K_\mu$ from the formula in the appendix, yielding:
\eqb
  (K_{i\rho}, K_{i\lambda})  
   &\to& \frac{\pi^2}{2 \rho \sinh \pi \rho} 
    [ \delta(\rho+\lambda) + \delta(\rho-\lambda) ]
  \period \label{KK}
\eqe
The function space can be decomposed as follows:
\eqb
  g(t) &=& \frac{2}{\pi^2} \int_{0}^\infty  
    C_\rho K_{i\rho}(t)  \, \rho \sinh \pi \rho \, d\rho  \comma \nn \\
  |g(t)|^2 &=& (g, g^*) \ = \ 
    \frac{2}{\pi^2} \int_{0}^\infty |C_\rho|^2
    \rho \sinh \pi \rho \, d\rho
  \comma
\eqe 
where
\eqb
  C_\rho &=& (g, K_{i\rho}) 
  \period
\eqe
This again agrees with \cite{Titchmarsh}.
These spaces of functions  describe the components of the Hilbert space associated with 
particles that have energy larger or smaller than their angular momentum respectively.
The double overlaps follow trivially from the above results.
\subsection{Three-point function}
We have clarified the structure of the Hilbert space, and we computed the double
overlaps in the previous section. 
The three-point functions in our minisuperspace limit in Lorentzian
signature
can be computed using known integrals of the product of
three Bessel functions. We discuss
the case for a triple overlap between wavefunctions in the 
continuous series in some detail. The other cases are interesting,
and analogous. We then move on to make the connection to the Clebsch-Gordan
coefficients of $SL(2,R)$ in a parabolic basis.
\subsubsection*{Correlation functions for the continuous series} 
The overlaps for the principal continuous series
with $2 \tau_i+1=\nu_i \in i R$
are given by
\eqb
  \langle \prod_{i=1}^N  K^{33} (\lambda_i,\mu_i;g;\chi_i) \rangle
 & = &
   \frac{1}{8 \pi^2} \int d v_r \int dv_l
    \int_{-\infty}^{+\infty} d \rho \, e^{2 \rho}
   \prod_{i=1}^N K^{33} (\lambda_i,\mu_i;g;\chi_i) \label{NK33} \\
 & = & 
   \delta\bigl(\Sigma (\mu_i + \lambda_i) \bigr) 
   \delta_{\Sigma (\mu_i -\lambda_i)}
\int_{-\infty}^{+\infty} d \rho \, e^{2 \rho}
\prod_{i=1}^N K^{33} (\lambda_i,\mu_i;\rho;\chi_i). \nn
\eqe 

Since $K^{33}$ are functions of $ x =e^{-\rho}$, we find that calculation
of the N-point functions reduces to integrals of the form
\eqb
   \int_0^\infty dx \, x^{N-3} \prod_{k=1}^N Z_{\pm \nu_i}(a_i x)
  \comma \label{Npt}
\eqe   
where $Z_{\nu_i} = J_{\nu_i} $ or $K_{\nu_i}$ $(\nu_i \in i R, a_i > 0)$. 
Let's concentrate on the three-point functions.
To proceed, we need the integrals (\ref{JJJ})-(\ref{KKK}) in the appendix,
with $\rho =1$.
Since $\nu_i \in iR, \, a_i >0 $ in (\ref{Npt}), 
we find that the conditions for the
formulas (\ref{condZZZ}) are always satisfied, except for $a_3>a_1+a_2$ for 
(\ref{JJJ}). 
However, we can choose $a_i$ so that 
$a_3 \geq a_1+a_2$ is satisfied. From the 
conservation of $\lambda_i, \mu_i$, we have
\eqb
   \lambda_1+ \lambda_2 + \lambda_3 = 0  \comma && \mu_1+\mu_1 +\mu_3 = 0  
\period
   \label{cons}
\eqe
Note that (\ref{JJJ}) is needed only when 
$\lambda_i \mu_i > 0$, and denote $a_i = 2\sqrt{\lambda_i \mu_i}$. Then we obtain:
\eqb
  a_3^2 - (a_1+a_2)^2 &=& 4 \lb \sqrt{\lambda_1 \mu_2} - \sqrt{\lambda_2 \mu_1}
      \rb^2 \period
\eqe
Also, from (\ref{cons}), two of $\lambda_i (\mu_i)$ have the same sign.
So, we choose $ \lambda_{1,2} (\mu_{1,2})$ so that they 
have the same sign and $a_3 \geq a_1+a_2$.
Moreover, we can set one of 
$\lambda_1 \mu_2$ and $\lambda_2 \mu_1$ is greater than or equal to the other. 
Thus, in the following we assume
\eqb
   a_3 > a_1+a_2 \comma  && \lambda_1 \mu_2 \geq \lambda_2 \mu_1
   \label{ineq}
\eqe
for (\ref{JJJ2}), without loss of generality. 

\subsubsection*{Three-point function of $K^{33}$ with $\lambda_a \mu_a > 0$}
We will compute the triple overlap of two kernels with a third,
complex conjugated kernel. We will schematically denote the overlap
$<KKK^{\ast}>$ below.   The signs of $\lambda_3$ and $\mu_3$ are
 flipped in (\ref{cons}) under complex conjugation. 
It will be convenient in this case to make
an even more restrictive choice for the angular momentum and energy.
We take all $\lambda_i,
\mu_i > 0$. All other cases can be treated analogously, although we don't
exhibit them in detail. 
Under these conditions, 
$z_\pm$ in (\ref{JJJ2}) become
\eqb
   z_+ = \frac{\lambda_1}{\lambda_3} \comma &&
   z_- = \frac{\mu_2}{\mu_3} \period
\eqe
We have that the parameter $\delta_i=0$ in the kernels. 
Here we first consider the overlap
for $SL(2,R)$ to facilitate the connection to the group theory discussed 
in the next subsection. This leads to summing over the two patches
with $l=0$ and $l=1$, and to the parity conservation law 
$ 2 \sum \epsilon_i \equiv 0 \, \mbox{mod} 2$. We then obtain:
\begin{eqnarray}
<KKK^{\ast}> &=& \frac{ i \delta(\lambda_1+\lambda_2-\lambda_3)
\delta(\mu_1+\mu_2-\mu_3)}{2^3 \sin \pi(\frac{\nu_1}{2}+\epsilon_1) 
\sin \pi(\frac{\nu_2}{2}+\epsilon_2)
 \sin \pi(\frac{-\nu_3}{2}+\epsilon_3)} \nonumber \\
& & \times \, (\frac{\mu_1}{\lambda_1})^{\frac{\nu_1}{2}}
(\frac{\mu_2}{\lambda_2})^{\frac{\nu_2}{2}}
(\frac{\lambda_3}{\mu_3})^{\frac{\nu_3}{2}} 
\sum_{\sigma_i = \pm 1} (-1)^{\Sigma \sigma_i (\epsilon_i + \frac{1}{2})} 
I_3 (\sigma_i \nu_i),
\end{eqnarray}
where $I_3$ denotes the integral of three Bessel functions 
$J_{\sigma_i \nu_i} (a_i x)$ given in the appendix in formula (\ref{JJJ2}).
We will reshape this result for the three-point overlap such that it will become clearer how it
relates to the Clebsch-Gordan coefficients of $SL(2,R)$ in a parabolic basis.
First of all, we remark that it is easy to sum the two contributions arising from
$\sigma_3=\pm1$. Next, we use the formula
\eqb
  \F(a,b;c;z) &=& (1-z)^{-a} \F(a,c-b;c; \frac{z}{z-1}) 
    \comma 
\eqe
to express the hypergeometric functions in terms 
of the arguments $-\frac{\lambda_1}{\lambda_2}$
and  $-\frac{\mu_2}{\mu_1}$. Thus, we find the following slightly more compact expression for
the three-point function:
\eqb
&& <KKK^{\ast}> =
\frac{i(-1)^{\epsilon_3+\frac{1}{2}} \delta(\lambda_1+\lambda_2-\lambda_3)
\delta(\mu_1+\mu_2-\mu_3)}{8 \pi \sin \pi(\frac{\nu_1}{2}+\epsilon_1)
 \sin \pi(\frac{\nu_2}{2}+\epsilon_2)}  
  \lambda_2^{\frac{-1-\nu_1-\nu_2+\nu_3}{2}} 
\mu_1^{\frac{-1+\nu_1-\nu_2-\nu_3}{2}} \mu_2^{\nu_2} \nonumber \\
& &  \quad \times \, 
   \sum_{\sigma_{1,2} = \pm 1} 
(-1)^{\sigma_1(\epsilon_1+\frac{1}{2}) + \sigma_2(\epsilon_2+\frac{1}{2})} 
(\frac{\lambda_2}{\lambda_1})^{\nu_1(1-\sigma_1)/2}
(\frac{\mu_1}{\mu_2})^{\nu_2(1-\sigma_2)/2}
\nonumber \\
& & \qquad   
\sin \frac{\pi}{2} (-\sigma_1 \nu_1 -\sigma_2 \nu_2+ 2 \epsilon_3)  
\G{\frac{1+\sigma_1 \nu_1+\sigma_2 \nu_2+\nu_3}{2}, 
\frac{1+\sigma_1 \nu_1+\sigma_2 \nu_2-\nu_3}{2} }{\sigma_1\nu_1+1, \sigma_2 \nu_2+1}
\nonumber \\
& & \qquad  
   \F ( 
      \frac{1+\sigma_1 \nu_1+\sigma_2 \nu_2-\nu_3}{2}, 
      \frac{1+\sigma_1 \nu_1-\sigma_2 \nu_2-\nu_3}{2};
      \sigma_1 \nu_1+1; - \frac{\lambda_1}{\lambda_2}  
       ) \nn \\
&& \qquad 
 \F ( 
     \frac{1+\sigma_1 \nu_1+\sigma_2 \nu_2+\nu_3}{2}, 
     \frac{1-\sigma_1 \nu_1+\sigma_2 \nu_2+\nu_3}{2};
     \sigma_2 \nu_2+1;  - \frac{\mu_2}{\mu_1}  
           )
   \comma  
\eqe
with 
$ \Gamma{a_1,a_2,... \brack b_1,b_2,...} 
= {\Gamma(a_1) \Gamma(a_2) ... \o \Gamma(b_1) \Gamma(b_2)...}$.
This is a rather complicated formula for the three-point function in the 
momentum basis. We will gain a lot of insight in it by rewriting it as
a product of Clebsch-Gordan coefficients. We have moulded it into a form that
will naturally arise from the group theory analysis in the next section. 
 For the unextended massless black hole, we have only to set 
$\epsilon_i = 0$ and replace the delta functions as in (\ref{NK33}). 
  
\subsection{Underlying group theory}
These three-point overlaps are related to the Clebsch-Gordan coefficients 
of $SL(2,R)$
in a parabolic basis. Unfortunately, these CGC's are not known 
in a parabolic basis,
so we need to derive them. That is a non-trivial exercise in group
theory. Results on CGC's in an elliptic basis, 
relevant for strings
in $AdS_3$, were known  (see e.g. \cite{CGC} and references therein).
 We follow some of the techniques developed in that context to derive
the CGC's in a parabolic basis. In particular, we follow closely the analysis in \cite{VKS}.
We start by considering homogeneous functions that carry an $SL(2,R)$
representation:
\begin{eqnarray}
(T_{\chi}(g) F)(x,y) &=& F(\alpha x + \gamma y, \beta x + \delta y).
\end{eqnarray}
The function $F$ satisfies:
\begin{eqnarray}
F(cx, cy) &=& |c|^{2 \tau} \mbox{sign}^{2 \epsilon} (c) F(x,y),
\end{eqnarray}
and we will characterize the representation by its quadratic
Casimir and parity $\chi=(\tau,\epsilon)$.
We can take the same representation to be defined on a function
$f$ on the line $(x,1)$ parametrized by $x$. (We will often denote the one
argument of $f$ by $x$, abusing notation. We hope that does not lead to
confusion with the first argument of $F(x,y)$.) 
The representation acts as:
\begin{eqnarray}
(T_{\chi}(g) f)(x) &=& |\beta x + \delta|^{2 \tau}  \mbox{sign}^{2 \epsilon}
(\beta x + \delta) f(\frac{\alpha x+ \gamma}{\beta x + \delta}).
\end{eqnarray}
We can define a kernel $K$ that intertwines the product of two representations
with a third:
\begin{eqnarray}
f(x_1; \chi_1) f(x_2; \chi_2) &=& \int K_{\chi_1 \chi_2}^{\chi_3} (x_1,x_2;x_3)
f(x_3; \chi_3) dx_3 d j_3. \label{defkernel}
\end{eqnarray}
The integral is over the space of irreducible representations (labelled
by $j_3$), and the quantum
numbers in those representations (labelled by $x_3$).
By acting on both sides of the equation with $T(g)$, we can find a relation between
the kernel and its transform.
We obtain:
\begin{eqnarray}
K(x_i) &=& K(\frac{\alpha x_i + \gamma}{ \beta x_i + \delta})
 |\beta x_1 + \delta|^{2 \tau_1}  \mbox{sign}^{2 \epsilon_1} 
(\beta x_1 + \delta)
|\beta x_2 + \delta|^{2 \tau_2}  \mbox{sign}^{2 \epsilon_2}(\beta x_2 + \delta)
\nonumber \\ & & 
 |\beta x_3 + \delta|^{-2 \tau_3-2}  \mbox{sign}^{2 \epsilon_3}(\beta x_3 + \delta).
\end{eqnarray}
Now we look for a solution to this equation of the form suggested by the
$SL(2,R)$ invariant $(x_i y_j-y_i x_j)$:
\begin{eqnarray}
K(x_i) &=& \kappa |x_1-x_2|^{-2 a_3} |x_2-x_3|^{- 2a_1} |x_3-x_1|^{-2a_2} \nonumber \\ & & 
 \mbox{sign}^{2 \eta_3}  (x_1 -x_2)\mbox{sign}^{2 \eta_1}  (x_2 -x_3)\mbox{sign}^{2 \eta_2}  (x_3 -x_1),
\end{eqnarray}
with $\eta_i = 0,1/2$.
We then find two solutions of this form parametrised by $\omega \equiv \sum \eta_i \, \mbox{mod} \, 1$. In the light of
the representation theory of $SL(2,R)$ it is no surprise that we find two solutions, since the product of
two continuous representations contains two copies of each irreducible continuous representation
(e.g. \cite{VK}).
 We also have $\sum 2 \epsilon_i \equiv 0 \, \mbox{mod} \, 2$ and
\begin{eqnarray}
2a_1 &=& \tau_1-\tau_2+\tau_3+1 \nonumber \\
2a_2 &=& -\tau_1+\tau_2+\tau_3+1 \nonumber \\
2a_3 &=& -\tau_1-\tau_2-\tau_3-1 \nonumber \\
2 \eta_i & \equiv & 2 \epsilon_i + 2 \omega \, (\mbox{mod} \, 2).
\end{eqnarray}
By expressing the relation (\ref{defkernel}) in an orthonormal basis
of functions, we can show that the parabolic CGC are related through an overall constant
to the following integral (see \cite{VK} for an alternative derivation):
\begin{eqnarray}
 C_\omega(\lambda) &=& \kappa \int_{-\infty}^{+\infty} dx_1 dx_2 dx_3
|x_2-x_3|^{-2a_1} sign^{2 \eta_1}(x_2-x_3)
|x_3-x_1|^{-2a_2} sign^{2 \eta_2}(x_3-x_1)
\nonumber \\
& & 
|x_1-x_2|^{-2a_3} sign^{2 \eta_3}(x_1-x_2)
e^{i(\lambda_1 x_1 + \lambda_2 x_2 - \lambda_3 x_3)}.
\end{eqnarray}
In the appendix it is shown that these Clebsch-Gordan coefficients can be
computed to be (for $\lambda_i >0$):
\begin{eqnarray}
&& C_{\omega} (\lambda) = 8 \kappa   \delta(\lambda_1+\lambda_2-\lambda_3)
\Gamma(\textstyle{ \frac{1-\nu_1+\nu_2-\nu_3}{2} })
\Gamma(\textstyle{ \frac{1+\nu_1-\nu_2-\nu_3}{2} })
\Gamma(\textstyle{ \frac{1+\nu_1+\nu_2+\nu_3}{2} })
\nonumber \\ & & \quad
\prod_i \sin(\pi( a_i- \eta_i)) (-i)^{2 \eta_i}
 \lambda_2^{\frac{-\nu_1-\nu_2+\nu_3-1}{2}} \times
\nonumber \\
& & \quad \Bigl\{ 
  \F(\textstyle{
   \frac{\nu_1-\nu_2-\nu_3+1}{2},\frac{\nu_1+\nu_2-\nu_3+1}{2}; \nu_1+1;
-\frac{\lambda_1}{\lambda_2} 
  })
\G{\frac{\nu_1-\nu_2+\nu_3+1}{2},\frac{\nu_1+\nu_2-\nu_3+1}{2}}{\nu_1+1}
\nonumber \\
& &  \quad 
((-1)^{2 \omega} - \frac{(-1)^{2 \epsilon_1} \cos \frac{\pi}{2} (\nu_1+\nu_2-\nu_3)+
 \cos \frac{\pi}{2} (\nu_1-\nu_2+\nu_3)}{\sin \pi \nu_1})
\nonumber \\
&& \quad + \, (\frac{\lambda_1}{\lambda_2})^{-\nu_1} 
 \F(\textstyle{
\frac{-\nu_1+\nu_2-\nu_3+1}{2},\frac{-\nu_1-\nu_2-\nu_3+1}{2}; -\nu_1+1;
-\frac{\lambda_1}{\lambda_2}
})
\G{\frac{-\nu_1+\nu_2+\nu_3+1}{2},\frac{-\nu_1-\nu_2-\nu_3+1}{2}}{-\nu_1+1}
\nonumber \\
& &  \quad 
((-1)^{2 \omega+2\epsilon_2} + \frac{(-1)^{2 \epsilon_1} \cos \frac{\pi}{2} (-\nu_1+\nu_2+\nu_3)+
 \cos \frac{\pi}{2} (\nu_1+\nu_2+\nu_3)}{\sin \pi \nu_1}) \Bigr\},
\end{eqnarray}

\subsection{Towards the three-point function}
To obtain the three-point function for continuous representations, we have
 taken into account the fact that
 there are two  Clebsch-Gordan coefficients
associated to each triple of continuous representations. These two 
CGC's were labelled by $\omega$. We can compute now the following combination
of the CGC's which does not depend on a choice of basis in each of these equivalent irreducible 
components in the tensor product representation:
\begin{eqnarray}
&& C_{\omega=0}(\lambda) C_{\omega=0}(\mu)^{\ast} 
  + C_{\omega=1/2}(\lambda) C_{\omega=1/2}(\mu)^{\ast}
 \nonumber \\ 
&& \quad = |\kappa|^2 (2\pi)^4
\frac{\delta(\lambda_1+\lambda_2-\lambda_3)
\delta(\mu_1+\mu_2-\mu_3)}{\sin \pi(\frac{\nu_1}{2}+\epsilon_1)
 \sin \pi(\frac{\nu_2}{2}+\epsilon_2)}  
  \lambda_2^{\frac{-1-\nu_1-\nu_2+\nu_3}{2}} 
\mu_1^{\frac{-1+\nu_1-\nu_2-\nu_3}{2}} \mu_2^{\nu_2} \nonumber \\
& &  \quad \quad \times \, 
   \sum_{\sigma_{1,2} = \pm 1} 
(-1)^{\frac{1}{2}(\sigma_1+\sigma_2)} 
(\frac{\lambda_2}{\lambda_1})^{\nu_1(1-\sigma_1)/2}
(\frac{\mu_1}{\mu_2})^{\nu_2(1-\sigma_2)/2}
\nonumber \\
& & \qquad \quad   
\sin \frac{\pi}{2} [ \sigma_1 (\nu_1+2\epsilon_1) 
   + \sigma_2 (\nu_2+2\epsilon_2)]  
 \G{\frac{1+\sigma_1 \nu_1+\sigma_2 \nu_2+\nu_3}{2}, 
\frac{1+\sigma_1 \nu_1+\sigma_2 \nu_2-\nu_3}{2} }{\sigma_1\nu_1+1, \sigma_2 \nu_2+1}
\nonumber \\
& & \qquad  \quad 
   \F ( { 
      \frac{1+\sigma_1 \nu_1+\sigma_2 \nu_2-\nu_3}{2}, 
      \frac{1+\sigma_1 \nu_1-\sigma_2 \nu_2-\nu_3}{2};
     \sigma_1 \nu_1+1; - \frac{\lambda_1}{\lambda_2}  
       }) \nn \\
&& \qquad \quad 
 \F ( { 
     \frac{1+\sigma_1 \nu_1+\sigma_2 \nu_2+\nu_3}{2}, 
     \frac{1-\sigma_1 \nu_1+\sigma_2 \nu_2+\nu_3}{2};
     \sigma_2 \nu_2+1;  - \frac{\mu_2}{\mu_1}  
         }  ) \period 
\end{eqnarray}
To calculate $C_{\omega}(\mu)^{\ast}$, we first of all made use of 
the symmetry of the integral $I'$ that we noted in the appendix,
to express the CGC as a function of the argument $-\frac{\mu_2}{\mu_1}$ 
(making use of the fact that the $\mu_i$ are positive by choice). 
To that end we need to exchange $\nu_{1}$ with $\nu_{2}$,  
and $\lambda_{1,2}$ with $\mu_{2,1}$, and multiply 
with an overall $(-1)^{2 \omega}$. 
Finally, to take the complex conjugate, we simply map $\nu_i$ to
$-\nu_i$ for these continuous representations.
After using (a lot of) standard trigonometric identities  and
comparing the coefficients of the hypergeometric functions,
the result we find agrees with 
 the computation of the triple overlap via the integral
over three Bessel functions, for the choice of normalisation constant $|\kappa|^2=
\frac{1}{(2 \pi)^3} \frac{1}{ (4 \pi)^2}$.

Of course, the three-point functions for other choices of representation, and energy and angular momentum
can similarly be related to Clebsch-Gordan coefficients. It would be interesting to explicitly
prove that statement, especially in the cases that involve discrete representations. The analysis
should be doable using our results and the example of known results in the elliptic basis.

\subsection{Four-point function}
For the four-point overlaps, we need the integral of the product of four Bessel functions.
These (along with integrals relevant for higher point-functions) 
can be found in \cite{Prudnikov}. We won't discuss the four-point function 
in detail here -- we only make a few comments.
The Fourier-transformed integral has been studied 
in the context of the $AdS/CFT$ correspondence. 
Even before making that statement
clearer in the next section, we already note that the analysis of the singular
parts of the four-point function in the case of $AdS_5/CFT_4$ (see e.g. \cite{D'Hoker:2002aw} for a review) gave rise to
not only power law singularities, but also logarithmic singularities that were interpreted \cite{Bianchi:1999ge}
as arising from operators with anomalous dimensions. It would certainly be interesting
to revisit the analysis of the singularity structure in the two-dimensional context,
to analyze factorisation of the four-point function (even at the level of the
minisuperspace limit). Of course, one would keep an eye out for the singularities
appearing in the four-point functions in ordinary (flat space) parabolic orbifolds \cite{Liu:2002kb},
and analyze whether these arises in the context of the massless BTZ black hole.
It would be interesting to see whether and how the instability mechanism pointed
out in \cite{Horowitz:2002mw} is at work here. 

\section{Holographic dual}
In this section, we want to show in what way our work is related to
previous work on the $AdS/CFT$ correspondence. One of the main differences
is that we worked in Lorentzian signature. Aspects of the correspondence
in Lorentzian signature have been studied in
 \cite{Son:2002sd}\cite{Balasubramanian:1998sn}\cite{Balasubramanian:1998de}\cite{Balasubramanian:1999ri}.
In particular the authors of \cite{Balasubramanian:1999ri} point out that it is natural to concentrate
on Poincare patches within $AdS$ space to formulate a Lorentzian version of the correspondence. We 
note that in the context of the massless BTZ black hole, this is very naturally realised 
(see subsection \ref{massless}).

In this section we review some holographic aspects of strings on BTZ black holes, and we add
a few comments on aspects that are particular to the extremal and massless black hole. We believe
this clarifies some open ends in the literature, and shows that our results can find applications
in a more general context.

\subsection{Bulk-to-boundary}
It is well known in the euclidean $AdS$ context that the
bulk-boundary propagator in Poincar\'e coordinates is:
\begin{eqnarray}
K_P (z,\lambda_+,\lambda_-) &=& c (\frac{z}{z^2+ \Delta \lambda_+ \Delta \lambda_-})^{2 h_+},
\end{eqnarray}
where $c=\frac{2 h_+ -1}{\pi}$ and $2 h_+$ is related to the scaling
 dimension of the boundary field.
The bulk field $\Phi$ can then be written 
in terms of the boundary condition $\Phi_0$:
\begin{eqnarray}
\Phi(z,\lambda_+,\lambda_-) &=& \int d \lambda_+' d \lambda_-' K_P(z,\lambda_+,\lambda_-;
\lambda_+',\lambda_-') \Phi_0 (\lambda_+',\lambda_-'). \label{bulksol}
\end{eqnarray}
In Lorentzian signature, there are normalisable modes in the bulk that allow for many
bulk solutions to the same boundary conditions. We assume, 
until subsection \ref{lorcor}, that these take the values that
follow from analytic continuation from the euclidean case, for simplicity.
We thus specify a particular coherent 
state in the boundary conformal field theory \cite{Balasubramanian:1998de}.

Of course, we know that the Poincare coordinates are very appropriate for a description
of the massless BTZ black hole, and we therefore needn't look further for the bulk-to-boundary
propagator for the massless BTZ black hole. Let us nevertheless follow a roundabout 
route to reobtain the propagator to illustrate some important points.

We can derive the bulk-to-boundary propagator for the generic BTZ black hole 
background by performing some coordinate transformations starting from 
a Poincare patch. We need to carefully take into 
account the transformation of the measure in  (\ref{bulksol}) as well
as the conformal transformation of the boundary field (with weights $h_-=
1-h_+$)
(see e.g. \cite{Keski-Vakkuri:1998nw}). The coordinate transformations to obtain the
generic BTZ black hole metric are:
\begin{eqnarray}
\lambda_{\pm} &=& \sqrt{\frac{r^2-r_+^2}{r^2-r_-^2}} e^{2 \pi T_{\pm} u_{\pm}} \nonumber \\
z &=&  \sqrt{\frac{r_+^2-r_-^2}{r^2-r_-^2}} e^{ \pi (T_{+} u_{+}+ T_- u_-)},
\label{poingen}
\end{eqnarray}
where $u_\pm = -\phi \pm t$.
A computation (with $\phi$ non-compact)
 yields the propagator\footnote{We evaluate
$ \frac{d\lambda'_+}{du'_+}$ at $r=\infty$ since $\Phi_0$ is 
a conformal field at  the boundary.}: 
\begin{eqnarray}
K_{BTZ} &=& c (4 \pi^2 T_+ T_-)^{2 h_+} (\frac{r^2-r_-^2}{(r^2-r_+^2)^2})^{h_+}
 e^{-2 \pi h_+ (T_+ \Delta
u_+ +T_- \Delta u_-)} \nonumber \\
&\times & (\frac{r_+^2-r_-^2}{r^2-r_+^2}+(1-e^{-2 \pi T_+ \Delta u_+})(1-e^{-2 \pi T_- \Delta u_-}))^{-2 h_+}.
\end{eqnarray}
Note that we did not make the approximation used in  \cite{Keski-Vakkuri:1998nw}, which explains for the
slightly different expression for the propagator.
Similarly, following the coordinate transformations in
\cite{Maldacena:1998bw} or our section \ref{global}, we  
obtain the bulk-boundary propagators for the extremal and
the massless BTZ black hole (for $\phi$ non-compact):
\begin{eqnarray}
K_{EBH} &=& c e^{-2 \pi h_+ T_+ \Delta u_+} (\frac{1}{r^2-r_+^2})^{h_+} 
 (\frac{1}{r^2-r_+^2}+\frac{1}{2 \pi T_+}(1-e^{-2 \pi T_+ \Delta u_+}) \Delta u_-)^{-2 h_+} \nonumber \\
K_{M=0} &=& c  (\frac{r}{1+r^2 \Delta u_+ \Delta u_-})^{2 h_+}.
\end{eqnarray}
It is a posteriori obvious that, although the coordinate transformation 
(\ref{poingen}) to go from the Poincare 
patch to the generic BTZ background degenerates for the extremal cases, the bulk-to-boundary
propagator for these extremal cases can be obtained straightforwardly by taking the
appropriate limit of the bulk-to-boundary propagator. Thus, we needn't be worried when taking limits
of the generic BTZ results based on the bulk-to-boundary propagator. 

As we anticipated, the bulk-to-boundary
propagator that we obtained for the massless black hole
is the Poincare bulk-to-boundary propagator.

Finally, to obtain the propagators taking into account the periodicity of $\phi$, we need to
sum the propagator over images \cite{Hemming:2002kd}:
\begin{eqnarray}
K_{BTZ} &=& \sum_{n=\infty}^{- \infty}
c (4 \pi^2 T_+ T_-)^{2 h_+} (\frac{r^2-r_-^2}{(r^2-r_+^2)^2})^{h_+} e^{-2 \pi h_+ (T_+ (\Delta
u_+ +2 \pi n)  +T_- (\Delta u_- +2 \pi n))} \nonumber \\
&\times & (\frac{r_+^2-r_-^2}{r^2-r_+^2}+
(1-e^{-2 \pi T_+ (\Delta u_+ + 2 \pi n)})(1-e^{-2 \pi T_- (\Delta u_-+ 2 \pi n)}))^{-2 h_+}.
\end{eqnarray}

\subsection{One-point function}
A first aspect of the holographic correspondence that we want to stress is the
following.
The one-point function for the energy momentum tensor in these backgrounds was
computed in  \cite{Hemming:2002kd} on the basis of the supergravity solution.
The one-point function has to agree with the vacuum expectation value of the boundary
energy momentum tensor \cite{Balasubramanian:1999re}. 
The vacuum expectation value of the energy momentum tensor in the boundary theory
was computed in \cite{Maldacena:1998bw}. The computation
was based on performing a Bogoliubov transformation on the Poincare vacuum, taking the vacuum
to a thermally excited state in the 
CFT, for left- and rightmovers in the generic BTZ background,
and for (say) leftmovers only for the extremal BTZ background.
It is a trivial but important observation that indeed these two computations agree.

 The second aspect we
stress here is that the Poincare vacuum does not undergo a Bogoliubov transformation
for the massless black hole. Indeed, the massless black hole is described, precisely
by the group element parametrized in Poincare coordinates. 
In other words, the Poincare
vacuum in the boundary CFT agrees with the vacuum 
for the massless BTZ black hole 
background (up to the global angular identification $\phi \sim 2\pi n$).
We stress this because the standard association of 
the results in Poincare
coordinates to the $AdS_3$ background might be cause of confusion 
in our context. 

\subsection{Two-point function}
Using the bulk-to-boundary propagator, we can compute two- and three-point functions in the
boundary conformal field theory in a supergravity approximation. 
The two-point functions for operators $O,O'$ in the boundary theory are (see e.g. 
 \cite{Hemming:2002kd}\cite{Keski-Vakkuri:1998nw}):
\begin{eqnarray}
&& \bra O O' \ket  \sim (4\pi^2T_+T_-)^{2h_+}
 \sum_{n=\infty}^{- \infty} \sinh^{-2 h_+}(\pi T_+
(\Delta u_+ + 2 \pi n))  \sinh^{-2 h_+}(\pi T_-
(\Delta u_- + 2 \pi n)), \nonumber \\
\end{eqnarray}
and, as remarked before, they allow smooth limits in the extremal cases.
It was argued in \cite{Hemming:2002kd} that these can be recuperated from the two-point function
in the full string theory. Indeed, in the $AdS_3$ context, it is known that the boundary
correlation functions are obtained by integrating worldsheet correlation functions over the
worldsheet.

\subsection{Three-point function}
\label{lorcor}
What we want to point out here is the relation to the computation of the three-point functions
in a supergravity limit in euclidean signature, and to show that our results can be used to define a Lorentzian generalisation. 
(Note that this remark also applies to the two-point overlaps with some care 
about normalisations.)
The computation of the three-point function in euclidean signature in a supergravity
approximation \cite{Freedman:1998tz} was done in Poincare coordinates, in coordinate
space\footnote{Stricly speaking, once again, we need to sum over images to take into
account the periodicity of the $\phi$-coordinate.}.
We would like to compare these boundary conformal field theory three-point functions
in a supergravity approximation with our results for the pointparticle limit
of the worldsheet three-point function.

First of all, we note the relation between the bulk-to-boundary propagator in coordinate and
momentum space.
Using integrals noted in the appendix (and, in particular, performing the angular integral first), 
we can prove the following two-dimensional Fourier-transform (i.e. Fourier transform on the boundary):
\begin{eqnarray}
\int e^{-ikx} dx \frac{\Delta-1}{\pi} (\frac{z_0}{z_0^2+|x|^2})^{\Delta}
&=& \frac{2^{-\nu+1}}{\Gamma(\nu)} z_0 k^{\nu} K_{\nu}(k z_0) 
\end{eqnarray}
where $\nu=\Delta-1$. That coincides with the {\em naive} propagator in  \cite{Freedman:1998tz}
(which has been argued to be ill-suited for computing the two-point function -- it misses out on
a crucial normalisation factor necessary for verification of the Ward identities).
Thus, the Fourier transform of the Poincare bulk-to-boundary
 propagator is the modified Bessel function.
It should be clear now that the three-point function $A_1(x_i)$
in \cite{Freedman:1998tz} is related, via FT on the boundary,
to the integral over three modified Bessel functions. 
More explicitly, we should have that
\begin{eqnarray}
A_1(k_i) &=& \frac{1}{(2 \pi)^3} \int dx_i A_1(x_i) e^{-i \sum k_i x_i} \nonumber \\
  &=&  \frac{-\delta(\sum k_i)}{2 \pi} \int \frac{dw_0}{w_0^3} \prod_{i=1}^3 K_{\Delta_i} (w_0,k_i) \nonumber \\
&=& \frac{-4\delta(\sum k_i)}{ \pi} \int dw_0 \prod_{i=1}^3
 (2^{\nu_i} \Gamma(\nu_i))^{-1} k_i^{\nu_i} K_{\nu_i} (k_i w_0).
\end{eqnarray}
The second line contains the propagator $K_{\Delta_i}$
in momentum space, the third line modified Bessel functions $K_{\nu_i}$. 
The last line is the integral over three modified Bessel functions that we
encountered as the triple overlap of certain particle wavefunctions in the
massless black hole background. 

In a general $AdS/CFT$ context, analysed in Poincare coordinates, in Lorentzian
signature, we are free to choose a different propagator, with extra terms proportional to 
ordinary Bessel functions \cite{Balasubramanian:1998de}. 
This corresponds to choosing a different coherent state in the boundary
conformal field theory. If we want to compute the three-point function in this Lorentzian conformal
field theory, we would evaluate the integrals over a product of three Bessel functions in the bulk.
Thus, the integrals over three Bessel functions that we provided 
in section 4 and the
appendix allow for a computation of the 
three-point functions in Lorentzian signature with any given choice 
of propagator.\footnote{In a general $AdS/CFT$ context, 
the indices $\nu_i$ do not satisfy (\ref{condZZZ}). 
One may then define the integral outside the region (\ref{condZZZ})
 by analytic continuation.} 

Note that in  momentum space, we only needed to restrict $\mu-\lambda$ 
to integers to implement the angular identification for the BTZ black hole,
 and did not have to deal with the infinite sum over images. 
This is a definite advantage of working in momentum space.

\section{Conclusions and discussion}
We analysed the minisuperspace limit of the $SL(2,R)$ Wess-Zumino-Witten model, in a scheme appropriate
for the description of string propagation on the massless BTZ black hole background. In particular,
we discussed the appropriate Hilbert space (and string spectrum),
 the group theoretic structure of the three-point functions, and we analyzed  some
features that are typical of the Lorentzian signature of the background. 

One goal to keep in mind is the connection by spectral flow between the holographic 
dual boundary conformal field
theory to the theory dual to the $AdS_3$ background. It would be very useful to work out these
correspondences in some detail, since it might shed light on the difficult issue of time-dependent
(and possibly singular or unstable) backgrounds in string theory. To that end, it would be instructive
to analyse further the boundary conformal algebra, including its supersymmetric extensions,
in a string theoretic context, and to connect it via spectral flow in the boundary theory
to the analysis performed
in the $AdS_3$ case. It is also important to work out the four-point functions in the particle model,
and to connect it to the full four-point function in the (euclidean) conformal field theory. That
should enable one to analyze whether the instability found in the parabolic orbifold persists in
this background, and whether there is a sensible interpretation of the instability in this context,
perhaps via the boundary conformal field theory. Note that the existence of the boundary conformal
field theory may be associated directly to the fact that the background is asymptotically $AdS_3$
(and not Minkowski space).

There are other issues that are important to address. One is the extension of our analysis to
other BTZ black hole backgrounds. That will involve a substantial amount of $SL(2,R)$ group theory.
It is important, we believe, to do at least part of the analysis in the Lorentzian signature
spacetime, to get a better understanding of time-dependent features of the $AdS/CFT$
correspondence, and to be able to extract time-dependent physics from our knowledge of the
euclidean $SL(2,C)/SU(2)$ CFT.
Moreover, it would be interesting to reinterpret our results on three-point functions in a more
general Lorentzian $AdS/CFT$ context.

\acknowledgments
We would like to thank  Dan Freedman and Kenji Mohri
for correspondence and discussions. Our
research was supported in part by
University of Tsukuba Research Projects and the U.S. Department of Energy
under cooperative research agreement \# DE-FC02-94ER40818.

\appendix

\section{Integrals}
\label{integrals}

\subsection{Fourier transforms}

We note a few useful integral formulas:
\begin{eqnarray}
 \int_0^{2 \pi} d \phi e^{i a \cos \phi} &=& 2 \pi J_0(a) \nonumber \\
 \int_0^{\infty} x^{\mu} J_{\nu}(ax) dx &=& 2^{\mu} a^{-\mu-1} \frac{\Gamma(\frac{1+\nu+\mu}{2})}{\Gamma(\frac{1+\nu-\mu}{2})}
 \nonumber \\
 \int_0^{\infty} x^{\nu+1} J_{\nu}(bx) (x^2+a^2)^{-\mu-1} dx &=& \frac{a^{\nu-\mu} b^{\mu}}{2^{\mu} \Gamma(\mu+1)} K_{\nu-\mu}(ab).
\end{eqnarray}
The second formula is valid when $ -\re (\nu)-1< \re (\mu)<\frac{1}{2}, a>0$ 
and the third when
$-1< \re (\nu)< \re (2 \mu + \frac{3}{2}), a,b>0 $.
We also have:
\begin{eqnarray}
|x|^{-2 \nu-2} &=& \frac{1}{\pi} 2^{-2 \nu-2} \frac{\Gamma(-\nu)}{\Gamma(\nu+1)} \int dk e^{i k.x} |k|^{2 \nu}
\end{eqnarray}
which agrees with the appendix of \cite{Freedman:1998tz}.
\subsection{Integrals of Bessel functions}
\eqb
   (J_\mu, J_\nu) &\equiv & 
   \int_0^\infty {dx \o x} J_\mu (a x) J_\nu(a x) 
   \ = \ \frac{2}{\pi} {1 \o \mu^2 -\nu^2} \sin {\pi \o 2} (\mu -\nu) 
  \comma 
\eqe
which is valid for 
\eqb
  \mbox{Re} \, (\mu + \nu ) > 0 \comma \qquad  a > 0 \period
\eqe
 Next we have (\cite{Prudnikov}, p. 385):
\eqb
  (K_\mu, K_\nu)_{b,\alpha} &\equiv & 
   \int_0^\infty dx \, x^{\alpha-1} K_\mu (b x) K_\nu(b x) 
    \label{intKK} \\ 
   & = & 
   {2^{\alpha -3} \o b^{\alpha} \Gamma(\alpha)}
   \Gamma(\frac{\alpha+\mu+\nu}{2}) \Gamma(\frac{\alpha+\mu-\nu}{2})
   \Gamma(\frac{\alpha-\mu+\nu}{2})\Gamma(\frac{\alpha-\mu-\nu}{2})
  \comma \nn 
\eqe
which is valid for 
\eqb
  \re b > 0 \comma \qquad  
  \re \alpha > | \re \mu | + | \re \nu|
  \period \label{condK}
\eqe
Here we list formulas of integrals of three Bessel functions.  
We use the notation $ \Gamma{a_1,a_2,... \brack b_1,b_2,...} 
= {\Gamma(a_1) \Gamma(a_2) ...   \o \Gamma(b_1) \Gamma(b_2)...}$.
\eqb
  && \int_0^\infty dx \, x^{\rho-1} J_\lambda(ax) J_\mu(bx) J_\nu(cx) 
   \label{JJJ} \\
  && \quad = {2^{\rho-1} a^\lambda b^\mu \o c^{\rho+\lambda +\mu}}
   \G{\frac{\rho + \lambda + \mu +\nu}{2}
    }{\lambda +1, \mu+1, 1-\frac{\rho+\lambda+\mu-\nu}{2} }    
     F_4( \textstyle{ 
      \frac{\rho+\lambda+\mu-\nu}{2},\frac{\rho+\lambda+\mu+\nu}{2};
     \lambda+1,\mu+1; \frac{a^2}{c^2},\frac{b^2}{c^2} }) \comma  \nn \\
  && \int_0^\infty dx \, x^{\rho-1} J_\lambda(ax) J_\mu(bx) K_\nu(cx)
   \label{JJK}  \\
  && \quad = {2^{\rho-2} a^\lambda b^\mu \o c^{\rho+\lambda+\mu}} 
           \G{\frac{\rho + \lambda + \mu +\nu}{2}, 
     \frac{\rho+\lambda+\mu-\nu}{2}}{\lambda +1, \mu+1}    
    F_4(\textstyle{ 
      \frac{\rho+\lambda+\mu-\nu}{2},\frac{\rho+\lambda+\mu+\nu}{2};
     \lambda+1,\mu+1; -\frac{a^2}{c^2},-\frac{b^2}{c^2} }) \comma \nn  \\
  && \int_0^\infty dx \, x^{\rho-1} J_\lambda(ax) K_\mu(bx) K_\nu(cx) 
     = {2^{\rho-3} a^{\lambda} \o c^{\rho+\lambda}\Gamma(\lambda +1)  } 
       [ A(\mu) + A(-\mu)] \comma  \\
  && \quad       
    A(\mu) = (\frac{b}{c})^\mu 
    \Gamma\Bigl[ \textstyle{ -\mu, \frac{\rho + \lambda + \mu +\nu}{2}, 
      \frac{\rho+\lambda+\mu-\nu}{2}} \Bigr] 
    F_4(\textstyle{ 
      \frac{\rho+\lambda+\mu-\nu}{2},\frac{\rho+\lambda+\mu+\nu}{2};
     \lambda+1,\mu+1; -\frac{a^2}{c^2},\frac{b^2}{c^2} }) \comma \nn  \\
  && \int_0^\infty dx \, x^{\rho-1} K_\lambda(ax) K_\mu(bx) K_\nu(cx) 
   = {2^{\rho-4} \o c^{\rho}} \sum_{\sigma_1,\sigma_2 = \pm 1}
     A(\sigma_1 \lambda,\sigma_2 \mu) 
       \comma \nn  \\
  &&\quad       
    A(\lambda,\mu) = (\frac{a}{c})^\lambda (\frac{b}{c})^\mu 
    \Gamma \Bigl[ 
 \textstyle{ -\mu, -\lambda, \frac{\rho + \lambda + \mu +\nu}{2}, 
       \frac{\rho+\lambda+\mu-\nu}{2}} \Bigr] \label{KKK} \\
 && \qquad \qquad \qquad \qquad \qquad \times \, 
    F_4( \textstyle{ 
    \frac{\rho+\lambda+\mu-\nu}{2},\frac{\rho+\lambda+\mu+\nu}{2};
     \lambda+1,\mu+1; \frac{a^2}{c^2},\frac{b^2}{c^2} }) \period \nn  
\eqe
They are found in p. 717 in \cite{Gradsteyn}, p. 717 in \cite{Gradsteyn},
      p. 399 in \cite{Prudnikov}, p. 403 in \cite{Prudnikov}, 
and valid for 
\eqb
  \mbox{(i)} && \re (\lambda+\mu+\nu+\rho) > 0 ;\quad \re \rho < \frac{5}{2} ;
    \quad a,b > 0 ; \quad c > a+b  \nn \\
  \mbox{(ii)}  && \re (\lambda+\mu+\rho) > | \re \nu | ; \quad 
      \re c > | \im a | + | \im b | \nn \\
  \mbox{(iii)}  && \re (\lambda + \rho) > | \re \mu |  + | \re \nu | ; \quad 
      \re (b+c) > | \im a |  \nn  \\
   \mbox{(iv)} && \re \rho > | \re \lambda | + | \re \mu |  + | \re \nu | ; 
    \quad 
      \re (a+b+c) > 0      \label{condZZZ}
\eqe  
respectively. The formulas  (\ref{JJK})-(\ref{KKK}) may be obtained from 
 the first formula (\ref{JJJ}) by using the definition of $K_\mu$ in
terms of $J_\mu$.
The first formula may be obtained by expanding the first two Bessel functions,
and computing the moments of the third Bessel functions term by term.

For $\rho = 1$,  there is a formula
(p. 228 in \cite{Prudnikov}),   
\eqb
   && \int_0^\infty dx  \, J_\lambda(ax) J_\mu(bx) J_\nu(cx) 
    = {a^\lambda b^\mu \o c^{1+\lambda+\mu}} 
        \G{ \frac{\lambda + \mu +\nu+1}{2} 
         }{\lambda +1, \mu+1, \frac{1+\nu -\lambda-\mu}{2}} \label{JJJ2}
   \\    
     && \qquad \times
     \F ( \textstyle{ 
   \frac{1+\lambda+\mu-\nu}{2},\frac{1+\lambda+\mu+\nu}{2};
     \lambda+1; z_+ }) 
   \times 
      \F (\textstyle{ 
  \frac{1+\lambda+\mu-\nu}{2},\frac{1+\lambda+\mu+\nu}{2};
     \mu+1; z_-})
   \comma  \nn 
\eqe
where
\eqb
  && \re (\lambda+\mu+\nu) > -1 ; \quad a,b > 0 ; \quad c > a+b ; \nn \\
  &&  2c^2 z_\pm = \pm a^2 \mp b^2 + c^2 + 
     \sqrt{(a^2+b^2-c^2)^2-4a^2b^2} \period  
\eqe
This agrees with the identity between $F_4$ and $ \F $ in \cite{Slater}.

\section{Parabolic CGC}
We rewrite the integrand for $C_{\omega}(\lambda)$ using the following Fourier
transform:
\begin{eqnarray}
|x_i-x_j|^{-2 a_k} sign^{2 \eta_k}(x_i-x_j) &=&
\frac{\cos(\frac{\pi}{2}(-2 a_k+1+2 \eta_k))}{\pi}
\Gamma(-2 a_k+1) \nonumber \\
& & 
\int_{-\infty}^{+\infty} (-i sgn \xi)^{2 \eta_k}
|\xi|^{2 a_k-1} e^{-i\xi(x_i-x_j)} d \xi.
\end{eqnarray}
After performing the integrals over $x_i$ that yield three delta-functions,
we obtain:
\begin{eqnarray}
C_{\omega}(\lambda) &=& 8 \kappa \delta (\lambda_1+\lambda_2-\lambda_3)
\prod_{i=1}^3
\Gamma(-2a_i+1) \sin(\frac{\pi}{2}(2 a_i-2 \eta_i)) (-i)^{2 \eta_i}
I'_{\lambda}
\end{eqnarray}
where
\begin{eqnarray}
 I'_{\lambda}  & = &    
\int_{-\infty}^{+\infty} d \xi \, 
 sign^{2 \eta_1}(\lambda_2+\xi)sign^{2 \eta_2}(-\lambda_1+\xi)
sign^{2 \eta_3}(\xi) \nn \\
  && \qquad \times \, 
|\lambda_2+\xi|^{2a_1-1} |-\lambda_1+\xi|^{2a_2-1}|\xi|^{2a_3-1}.
\end{eqnarray}
(Note at this point that the integral is invariant under the 
exchange of $\lambda_{1,2}$, $\eta_{1,2}$ and $a_{1,2}$ up to
a factor of $(-1)^{2 \sum \eta_i}$.
We evaluate the integral $I'_{\lambda}$ in the case where $-\lambda_2<0<\lambda_1$, 
by dividing the real line into four regions, separated by the
points $-\lambda_2, 0, \lambda_1$. The four regions give rise to the
four terms:
\begin{eqnarray}
 I'_{\lambda} &=& 
(-1)^{2 \eta_1+2 \eta_2+2 \eta_3} \lambda_2^{2 a_1+2 a_2 + 2 a_3-2} 
B(-2 a_1-2 a_2 - 2 a_3+2, 2 a_1) \nonumber \\
& & \quad \times \, 
 \F (-2a_2+1,-2a_1-2a_2-2a_3+2; 
-2a_2-2a_3+2; -\frac{\lambda_1}{\lambda_2}) \nonumber \\
&+& 
(-1)^{2 \eta_2+2 \eta_3} \lambda_1^{2 a_2 -1}
\lambda_2^{2a_1+2a_3-1} 
B(2 a_1 , 2 a_3) \F (-2a_2+1,2a_3; 
2a_1+2a_3; -\frac{\lambda_2}{\lambda_1}) \nonumber \\
&+& 
(-1)^{2 \eta_2} \lambda_2^{2 a_1 -1}
\lambda_1^{2a_2+2a_3-1} 
B(2 a_2 , 2 a_3)  \F (-2a_1+1,2a_3; 
2a_2+2a_3; -\frac{\lambda_1}{\lambda_2}) \nonumber \\
&+&
 \lambda_1^{2 a_1+2 a_2 + 2 a_3-2} 
B(-2 a_1-2 a_2 - 2 a_3+2, 2 a_2) 
\nonumber \\
& & \quad \times \, \F (-2a_1+1,-2a_1-2a_2-2a_3+2; 
-2a_1-2a_3+2; -\frac{\lambda_2}{\lambda_1}).
\end{eqnarray}
Then we use the following relation to rewrite the expression in terms
of hypergeometric functions with the argument $-\frac{\lambda_1}{\lambda_2})$:
\begin{eqnarray}
&& (-z)^{-a} \F(a,a+1-c;a+1-b;z^{-1}) =  
\G{1-c,a+1-b}{1-b,a+1-c} \F(a,b;c;z)  \\
&& \qquad \qquad - \G{c,1-c,a+1-b}{2-c,c-b,a} e^{i \pi(c-1)} z^{1-c}
\F(1+a-c,b+1-c;2-c;z). \nonumber 
\end{eqnarray}
The resulting expression, after using  $\Gamma$-function
identities is:
\begin{eqnarray}
 I'_{\lambda} &=& \lambda_2^{2a_1+2a_2+2a_3-2} \times
\nonumber \\
& &  \Bigl\{ \F(-2a_2 \!+\! 1,-2a_1\!-\!2a_2\!-\!2a_3\!+\! 2; 
-2a_2\!-\!2a_3\!+\! 2;
-\frac{\lambda_1}{\lambda_2})
\G{2a_1,-2a_1\!-\!2a_2\!-\!2a_3\!+\! 2}{-2a_2\!-\!2a_3\!+\! 2}
\nonumber \\
& &  
\frac{(-1)^{2 \eta_1 + 2 \eta_2+ 2 \eta_3}\sin \pi (2a_2+2a_3)+
(-1)^{2 \eta_2+2 \eta_3}\sin \pi (2a_1+2a_2+2a_3)-
 \sin \pi 2 a_1}{\sin \pi (2a_2+2a_3)}
\nonumber \\
&+&  \, (\frac{\lambda_1}{\lambda_2})^{2a_2+2a_3-1} 
  \F(-2a_1+1,2a_3; 2a_2+2a_3;
-\frac{\lambda_1}{\lambda_2})
\G{2a_2,2a_3}{2a_2+2a_3}
\nonumber \\
& & 
\frac{(-1)^{2 \eta_2}\sin \pi (2a_2+2a_3)+
(-1)^{2 \eta_2+2 \eta_3}\sin \pi 2a_2+
 \sin \pi 2 a_3}{\sin \pi (2a_2+2a_3)}
\Bigr\}.
\end{eqnarray}

Now that we evaluated $I'_{\lambda}$, we can evaluate our original $C_{\omega}(\lambda)$, the
 CGC's in the parabolic basis. We plug in the values for the $a_i$ (in some instances) and
$\eta_i$, and rewrite the expression a little to obtain ($2 \tau_i+1=\nu_i$):
\begin{eqnarray}
&& C_{\omega} (\lambda) = 8 \kappa   \delta(\lambda_1+\lambda_2-\lambda_3)
\Gamma(\textstyle{ \frac{1-\nu_1+\nu_2-\nu_3}{2} })
\Gamma(\textstyle{ \frac{1+\nu_1-\nu_2-\nu_3}{2} })
\Gamma(\textstyle{ \frac{1+\nu_1+\nu_2+\nu_3}{2} })
\nonumber \\ & & \quad
\prod_i \sin(\pi( a_i- \eta_i)) (-i)^{2 \eta_i}
 \lambda_2^{\frac{-\nu_1-\nu_2+\nu_3-1}{2}} \times
\nonumber \\
& & \quad \Bigl\{ 
  \F(\textstyle{
   \frac{\nu_1-\nu_2-\nu_3+1}{2},\frac{\nu_1+\nu_2-\nu_3+1}{2}; \nu_1+1;
-\frac{\lambda_1}{\lambda_2} 
  })
 \G{\frac{\nu_1-\nu_2+\nu_3+1}{2},\frac{\nu_1+\nu_2-\nu_3+1}{2}}{\nu_1+1}
\nonumber \\
& &  \quad 
((-1)^{2 \omega} - \frac{(-1)^{2 \epsilon_1} \cos \frac{\pi}{2} (\nu_1+\nu_2-\nu_3)+
 \cos \frac{\pi}{2} (\nu_1-\nu_2+\nu_3)}{\sin \pi \nu_1})
\nonumber \\
&& \quad + \, (\frac{\lambda_1}{\lambda_2})^{-\nu_1} 
 \F(\textstyle{
\frac{-\nu_1+\nu_2-\nu_3+1}{2},\frac{-\nu_1-\nu_2-\nu_3+1}{2}; -\nu_1+1;
-\frac{\lambda_1}{\lambda_2}
})
\G{\frac{-\nu_1+\nu_2+\nu_3+1}{2},\frac{-\nu_1-\nu_2-\nu_3+1}{2}}{-\nu_1+1}
\nonumber \\
& &  \quad 
((-1)^{2 \omega+2\epsilon_2} + \frac{(-1)^{2 \epsilon_1} \cos \frac{\pi}{2} (-\nu_1+\nu_2+\nu_3)+
 \cos \frac{\pi}{2} (\nu_1+\nu_2+\nu_3)}{\sin \pi \nu_1}) \Bigr\},
\end{eqnarray}
which can be further simplified a  bit using trigonometry.

\end{document}